\documentclass{aastex}          
\usepackage{spr-astr-addons}    

\begin{document}
%
\title{The long-term survival chances of young massive star clusters}

\shorttitle{Survival of young massive star clusters}
\shortauthors{Richard de Grijs}

\author{Richard de Grijs\altaffilmark{1,2}} 
\affil{$^1$ Department of Physics \& Astronomy, The University of
Sheffield, Hicks Building, Hounsfield Road, Sheffield S3 7RH, U.K.}
\affil{$^2$ National Astronomical Observatories, Chinese Academy
of Sciences, 20A Datun Road, Chaoyang District, Beijing 100012, China}
\email{R.deGrijs@sheffield.ac.uk} 


\begin{abstract}
I review the long-term survival chances of young massive star clusters
(YMCs), hallmarks of intense starburst episodes often associated with
violent galaxy interactions. In particular, I address the key question
as to whether at least some of these YMCs can be considered
proto-globular clusters (GCs). In the absence of significant external
perturbations, the key factor determining a cluster's long-term
survival chances is the shape of its stellar initial mass function. I
conclude that there is an increasing body of evidence that GC
formation appears to be continuing until today; their long-term
evolution crucially depends on their environmental conditions,
however.
\end{abstract}


%

\section{Young mass star clusters as proto-globulars}
\label{intro.sec}

Young, massive star clusters (YMCs) are the hallmarks of violent
star-forming episodes triggered by galaxy collisions and close
encounters. Their contribution to the total luminosity induced by such
extreme conditions completely dominates the overall energy output due
to the interaction-induced star formation (e.g., de Grijs \&
Parmentier 2007; and references therein).

The question remains, however, whether or not at least a fraction of
the compact YMCs seen in abundance in extragalactic starbursts, are
potentially the progenitors of ($\ga 10$ Gyr) old globular cluster
(GC)-type objects -- although of higher metallicity than the
present-day GCs. If we could settle this issue convincingly, one way
or the other, such a result would have far-reaching implications for a
wide range of astrophysical questions, including our understanding of
the process of galaxy formation and assembly, and the process and
conditions required for star (cluster) formation. Because of the lack
of a statistically significant sample of YMCs in the Local Group,
however, we need to resort to either statistical arguments or to the
painstaking approach of case-by-case studies of individual objects in
more distant galaxies.

\subsection{The stellar initial mass function}

The evolution to old age of young clusters depends crucially on their
stellar initial mass function (IMF). If the IMF slope is too shallow,
i.e., if the clusters are significantly deficient in low-mass stars
compared to, e.g., the solar neighbourhood, they will likely disperse
within about a Gyr of their formation (e.g., Chernoff \& Shapiro 1987;
Chernoff \& Weinberg 1990; Goodwin 1997b; Smith \& Gallagher 2001;
Mengel et al. 2002). As a case in point, Goodwin (1997b) simulated the
evolution of $\sim 10^4 - 10^5$ M$_\odot$ YMCs similar to those observed
in the LMC, with IMF slopes $\alpha = 2.35$ (Salpeter 1955; where the
IMF is characterised as $\phi(m_\ast) \propto m_\ast^{-\alpha}$, as a
function of stellar mass, $m_\ast$) and $\alpha = 1.50$, i.e., roughly
covering the range of (present-day) mass function slopes observed in
LMC clusters at the time he performed his {\it N}-body simulations
(see also de Grijs et al. 2002a,b). The stellar mass range covered
ranged from 0.15 to 15 M$_\odot$; his {\it N}-body runs spanned at
most a few 100 Myr. Following Chernoff \& Weinberg (1990), and based
on a detailed comparison between the initial conditions for the LMC
YMCs derived in Goodwin (1997b) and the survival chances of massive
star clusters in a Milky Way-type gravitational potential (Goodwin
1997a), Goodwin (1997b; see also Takahashi \& Portegies Zwart 2000,
their fig. 8) concluded that -- for Galactocentric distances $\ga 12$
kpc -- some of his simulated LMC YMCs should be capable of surviving
for a Hubble time if $\alpha \ge 2$ (or even $\ga 3$; Mengel et
al. 2002), but not for shallower IMF slopes for any reasonable initial
conditions (cf. Chernoff \& Shapiro 1987; Chernoff \& Weinberg
1990). More specifically, Chernoff \& Weinberg (1990) and Takahashi \&
Portegies Zwart (2000), based on numerical cluster simulations
employing the Fokker-Planck approximation, suggest that the most
likely survivors to old age are, additionally, characterised by King
model concentrations, $c \ga 1.0-1.5$. Mengel et al. (2002; their
fig. 9) use these considerations to argue that their sample of YMCs
observed in the Antennae interacting system might survive for at least
a few Gyr, but see de Grijs et al. (2005), and Bastian \& Goodwin
(2006) and Goodwin \& Bastian (2006), for counterarguments related to
environmental effects and to variations in the clusters' (effective)
star-formation efficiencies (SFEs), respectively.

In addition, YMCs are subject to a variety of additional internal and
external drivers of cluster disruption. These include internal
two-body relaxation effects, the nature of the stellar velocity
distribution function, the effects of stellar mass segregation, disk
and bulge shocking, and tidal truncation (e.g., Chernoff \& Shapiro
1987; Gnedin \& Ostriker 1997). All of these act in tandem to
accelerate cluster expansion, thus leading to cluster dissolution --
since expansion will lead to greater vulnerability to tidally-induced
mass loss.

\subsection{Survival Diagnostics: the Mass-to-Light Ratio versus Age 
Diagram}

\begin{figure}[h!]
   \plotone{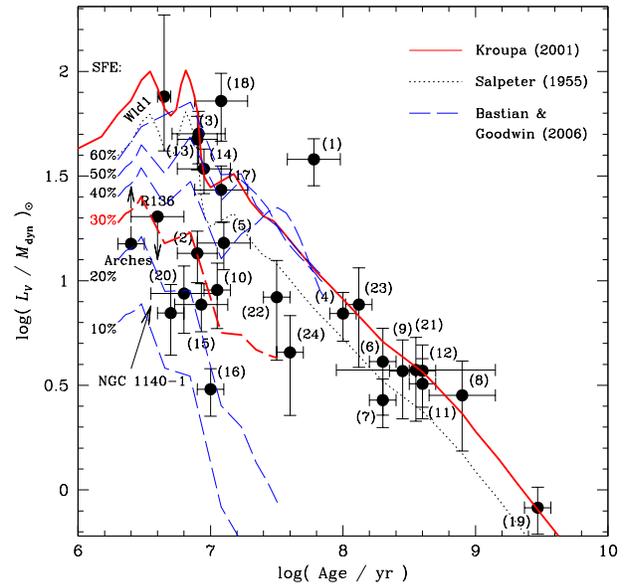}
   \caption{Updated version of the YMC $M/L$ ratio versus age
   diagnostic diagram. The numbered data points were taken from
   Bastian et al. (2006; and references therein); overplotted are the
   SSP predictions for a Salpeter (1955) and a Kroupa (2001) stellar
   IMF. We have included four new YMCs, NGC 1140-1 (Moll et al. 2007)
   the Galactic Centre Arches cluster, R136 in 30 Doradus (Goodwin \&
   Bastian 2006) and Westerlund 1 (denoted `Wld 1'; see de Grijs \&
   Parmentier 2007). The evolution expected for SSPs governed by IMFs
   as defined by both Salpeter (1955) and Kroupa (2001) is shown as
   the dolid and short-dashed lines, respectively. The long-dashed
   lines represent the evolution expected for SSPs with a Kroupa
   (2001)-type IMF, but a range of effective star-formation
   efficiencies (Goodwin \& Bastian 2006) }
   \label{diagnostic.fig}
\end{figure}

With the ever increasing number of large-aperture ground-based
telescopes equipped with state-of-the-art high-resolution
spectrographs and the wealth of observational data provided by the
{\sl Hubble Space Telescope}, we may now finally be getting close to
resolving the issue of potential YMC longevity conclusively. To do so,
one needs to obtain (i) high-resolution spectroscopy, in order to
obtain dynamical mass estimates, and (ii) high-resolution imaging to
measure their sizes (and luminosities). As a simple first approach,
one could then construct diagnostic diagrams of YMC mass-to-light
($M/L$) ratio versus age, and compare the YMC loci in this diagram
with simple stellar population (SSP) models using a variety of IMF
descriptions (cf. Smith \& Gallagher 2001; Mengel et al. 2002; Bastian
et al. 2006; Goodwin \& Bastian 2006). In Fig. \ref{diagnostic.fig} I
present an updated version of the $M/L$ ratio versus age diagram,
including all of the YMCs for which the required observables are
presently available (see also de Grijs \& Parmentier 2007). However,
such an approach, while instructive, has serious shortcomings. The
viability of this approach depends, in essence, on the validity of the
virial equation to convert line-of-sight velocity dispersions,
$\sigma_{\rm los}$, to dynamical mass estimates, $M_{\rm dyn}$, via
(Spitzer 1987):

\begin{equation}
\label{virial.eq}
M_{\rm dyn} = \frac{\eta \sigma^2_{\rm los} r_{\rm h}}{G} \quad ,
\end{equation}
where $r_{\rm h} = 1.3 \; R_{\rm eff}$ are the half-mass and effective
(or half-light) radii of the cluster, respectively, and $\eta = 3 a$;
$a \approx 2.5$ is the factor required to convert the half-mass to the
gravitational radius, $r_{\rm g}$. More specifically, following Fleck
et al. (2006), we write
\begin{equation}
\label{gravrad.eq}
r_{\rm g} = \frac{5}{2} \times \frac{4}{3} r_{\rm h} \quad ,
\end{equation}
where the factor 5/2 provides an approximate conversion for a large
range of clusters characterised by King (1966) mass profiles; the
second numerical factor in Eq. (\ref{gravrad.eq}) results from
projection on the sky, assuming that light traces mass throughout the
cluster. The use of both Eq. (\ref{virial.eq}) and the $M/L$ ratio
versus age diagram rely on a number of assumptions and degeneracies,
however, which I will discuss in some detail below.

\subsubsection{IMF degeneracies}

In the simplest approach, in which one compares the YMC loci in the
$M/L$ ratio versus age diagram with SSP models, the data can be
described by {\it both} variations in the IMF slope {\it and}
variations in a possible low-mass cut-off (e.g., Sternberg 1998; Smith
\& Gallagher 2001; Mengel et al. 2002); the models are fundamentally
degenerate for these parameters. For instance, Sternberg (1998)
derived for the YMC NGC 1705-I that it must either have a flat mass
function ($\alpha < 2$) or a low-mass truncation between 1 and 3
M$\odot$ (see also Smith \& Gallagher 2001); in both cases, it is
unlikely that this cluster may be capable of surviving for a Hubble
time.

However, the conclusion that the IMFs of such starburst clusters may
be unusual must be regarded with caution. As Smith \& Gallagher (2001)
point out, previous claims for highly abnormal (initial) mass
functions have often proven incorrect. If anything, the shape of the
mass function may vary on the size scales of the individual clusters,
but once one considers their birth environments on larger scales the
present-day mass function appears to be remarkably robust (e.g., Scalo
1998; Kroupa 2001), with the possible exception of the resolved
starburst clusters in the Milky Way (e.g., Stolte et al. 2005, 2006),
NGC 3603 and -- in particular -- the Galactic Centre Arches cluster.

Despite this controversy (particularly for some of the youngest
clusters), it appears that most of the YMCs for which high-resolution
spectroscopy is available are characterised by `standard' Salpeter
(1955) or Kroupa (2001) IMFs (e.g., Larsen et al. 2001; McCrady,
Gilbert \& Graham 2003; Maraston et al. 2004; Larsen, Brodie \& Hunter
2004; Larsen \& Richtler 2004; Bastian et al. 2006; see also de Grijs
et al. [2005] for a comparison of dynamical and photometric masses,
the latter based on `standard' IMF representations).

\subsubsection{Mass segregation}

While the assumption that these objects are {\it approximately} in
virial equilibrium is probably justified at ages greater than a few
$\times 10^7$ yr and for realistic SFEs $\ga 30$ per cent (at least
for the stars dominating the light; see, e.g., Goodwin \& Bastian
2006), the {\it central} velocity dispersion (as derived from
luminosity-weighted high-resolution spectroscopy) does not necessarily
represent a YMC's total mass. It is now well-established that almost
every YMC exhibits significant mass segregation from very young ages
onwards, so that the effects of mass segregation must be taken into
account when converting central velocity dispersions into dynamical
mass estimates (see also Fleck et al. 2006; Moll et al. 2007).

By ignoring the effects of mass segregation, as is in essence done if
one simply applies Eq. (\ref{virial.eq}), the underlying assumption is
then that of an isotropic stellar velocity distribution, i.e.,
$\sigma^2_{\rm total} = 3 \sigma^2_{\rm los}$, where $\sigma^2_{\rm
total}$ is the cluster's mean three-dimensional velocity dispersion.
In the presence of (significant) mass segregation in a cluster, the
central velocity dispersion will be dominated by the higher-mass stars
populating the cluster core. If we focus on dynamical evolution as the
dominant cause of mass segregation in clusters (as opposed to the
possible preferential formation of the higher-mass stars close to the
cluster core, also known as `primordial' mass segregation; e.g.,
Bonnell \& Davies 1998; de Grijs et al. 2002a), it follows that for
the high-mass stars to migrate to the cluster core, i.e., to the
bottom of the gravitational potential well, they must have exchanged
some of their kinetic energy with their lower-mass counterparts on
more extended orbits. As a consequence, the velocity dispersion
dominating the observed high-resolution spectra will be {\it lower}
than expected for a non-mass-segregated cluster of the same mass. In
addition, measurements of $r_{\rm h}$ will also be biased to smaller
values, and not to the values associated with the cluster as a
whole. Mass segregation will thus lead to an {\it under}\,estimate of
the true cluster mass.

I also note that the assumption of virial equilibrium only holds to a
limited extent, even in old GCs, because cluster-wide relaxation
time-scales of massive GC-type objects are of order $10^9$ yr or
longer (Djorgovski 1993). In fact, full global, or even local, energy
equipartition among stars covering a range of masses is never reached
in a realistic star cluster, not even among the most massive species
(e.g., Inagaki \& Saslaw 1985; Hunter et al. 1995). As the dynamical
evolution of a cluster progresses, low-mass stars will, on average,
attain larger orbits than the cluster's higher-mass stars, and the
low-mass stars will thus spend most of their time in the cluster's
outer regions, at the extremes of their orbits. For this reason alone,
we would not expect to achieve global energy equipartition in a
cluster. 

The time-scale for the onset of significant dynamical mass segregation
is comparable to the cluster's dynamical relaxation time (Spitzer \&
Shull 1975; Inagaki \& Saslaw 1985; Bonnell \& Davies 1998; Elson et
al.  1998). A cluster's characteristic time-scale may be taken as its
half-mass (or median) relaxation time, i.e., the relaxation time at
the mean density for the inner half of the cluster mass for cluster
stars with stellar velocity dispersions characteristic for the cluster
as a whole (Spitzer \& Hart 1971; Lightman \& Shapiro 1978; Meylan
1987; Malumuth \& Heap 1994).

Although the half-mass relaxation time characterises the dynamical
evolution of a cluster as a whole, significant differences are
expected locally within the cluster. The relaxation time-scale will be
shorter for higher-mass stars than for their lower-mass companions;
numerical simulations of realistic clusters confirm this picture
(e.g., Aarseth \& Heggie 1998; Kim et al. 2000; Portegies Zwart et
al. 2002). From this argument it follows that dynamical mass
segregation will also be most rapid where the local relaxation time is
shortest, i.e., near the cluster centre (cf. Fischer et al. 1998;
Hillenbrand \& Hartmann 1998). Thus, significant mass segregation
among the most massive stars in the cluster core occurs on the local,
central relaxation time-scale (comparable to just a few crossing
times; cf. Bonnell \& Davies 1998).

The combination of these effects will lead to an increase of the
dimensionless parameter $\eta$ in Eq. (\ref{virial.eq}) with time, if
the characteristic two-body relaxation time of a given (massive)
stellar species is short (Boily et al. 2005; Fleck et al. 2006), and
thus to an {\it under}\,estimate of the true cluster mass. However, we
note that Goodwin \& Bastian (2006) point out that a large fraction of
the youngest clusters in the $M/L$ ratio versus age diagram appear to
have dynamical masses well in excess of their photometric masses, and
that, therefore, the result of Boily et al. (2005) and Fleck et
al. (2006) does not seem applicable to these YMCs.

\subsubsection{Stellar masses}

Estimating dynamical masses, $M_{\rm cl}$, via Eq. (\ref{virial.eq})
assumes, in essence, that all stars in the cluster are of equal
mass. This is clearly a gross oversimplification, which has serious
consequences for the resulting mass estimates. The straightforward
application of the virial theorem tends to {\it under}\,estimate a
system's dynamical mass by a factor of $\sim 2$ compared to more
realistic multi-mass models (e.g., Mandushev, Spassova \& Staneva
1991; based on an analysis of the observational
uncertainties). Specifically, Mandushev et al. (1991) find that the
mass-luminosity relation for GCs with mass determinations based on
multi-component King-Michie models (obtained from the literature) lies
parallel to that for single-mass King models, but offset by $\Delta
\log M_{\rm cl} ({\rm M}_\odot) \simeq 0.3$ towards higher
masses. Farouki \& Salpeter (1982) already pointed out that cluster
relaxation and its tendency towards stellar energy equipartition is
accelerated as the stellar mass spectrum is widened; mass segregation
will then take place on shorter time-scales than for single-component
(equal-mass) clusters, and thus this will once again lead to an {\it
under}\,estimate of the true cluster mass (see also Goodwin 1997a;
Boily et al. 2005; Fleck et al. 2006; Kouwenhoven \& de Grijs 2008,
for multi-mass {\it N}-body approaches).

We also point out that if the cluster contains a significant fraction
of primordial binary and multiple systems, these will act to
effectively broaden the mass range and thus also speed up the
dynamical evolution of the cluster (e.g., Fleck et al. 2006;
Kouwenhoven \& de Grijs 2008).

\section{Cluster disruption at early times}
\label{disr.sec}

The early evolution of the star cluster population in the Small
Magellanic Cloud (SMC) has been the subject of considerable recent
attention and vigorous debate (e.g., Rafelski \& Zaritsky 2005;
Chandar, Fall \& Whitmore 2006; Chiosi et al. 2006; Gieles, Lamers \&
Portegies Zwart 2007). The key issue of contention is whether the
SMC's star cluster system has been subject to the significant early
cluster disruption processes observed in `normal', interacting and
starburst galaxies commonly referred to as `infant mortality' (e.g.,
Lada \& Lada 2003; Whitmore 2004; Bastian et al. 2005; Fall, Chandar
\& Whitmore 2005; Mengel et al. 2005; see also Whitmore, Chandar \&
Fall 2007) and `infant weight loss'. Chandar et al. (2006) argue that
the SMC has been losing up to 90 per cent of its star clusters per
decade of age, at least for ages from $\sim 10^7$ up to $\sim 10^9$
yr, whereas Gieles et al. (2007) conclude that there is no such
evidence for a rapid decline in the cluster population, and that the
decreasing number of clusters with increasing age is simply caused by
fading of their stellar populations. They contend that the difference
between their results was due to Chandar et al. (2006) assuming that
they were dealing with a mass-limited sample, whereas it is actually
magnitude-limited. In fact, this is not entirely correct; Chandar et
al. (2006) analyse the full magnitude-limited sample and conclude that
it is approximately surface-brightness limited. They then compare the
cluster age distribution of the full sample (expressed in units of
${\rm d}N_{\rm cl} / {\rm d}t$, i.e., the number of clusters per unit
time period) to that of a subsample for masses $\ge 10^3$ M$_\odot$
(which they do not analyse in the same manner), and suggest both to be
similar, although the latter is much flatter, hence giving rise to the
discrepancy between their results and those of Gieles et
al. (2007). Both studies are based on the same data set, the
Magellanic Clouds Photometric Survey (MCPS; Zaritsky, Harris \&
Thompson 1997).

\begin{figure}[t]
\includegraphics[width=\columnwidth]{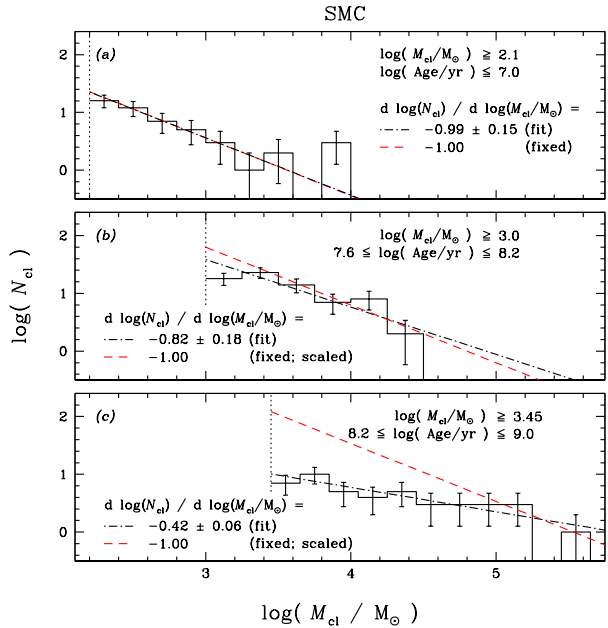}
\caption{CMFs for statistically complete SMC cluster subsamples. Age
and mass ranges are indicated in the panel legends; the vertical
dotted lines indicate the lower mass (50 per cent completeness) limits
adopted. Error bars represent simple Poissonian errors, while the
dashed lines represent CMFs of slope $\alpha = 2$, shifted vertically
as described in the text. The dash-dotted lines represent the best-fit
CMFs over the relevant mass range (see de Grijs \& Goodwin 2007)}
\label{smcclfs4.fig}
\end{figure}

In de Grijs \& Goodwin (2007), we set out to shed light on this
controversy surrounding the early evolution and disruption of star
clusters in the SMC. We embarked on a fresh approach to the problem,
using an independent, homogeneous data set of $UBVR$ imaging
observations, from which we obtained the cluster age distribution in a
self-consistent manner. We present the cluster mass functions (CMFs;
i.e., the number of clusters per unit mass range) for subsets of our
SMC cluster sample in Fig. \ref{smcclfs4.fig}, where the cluster
subsamples were selected based on their age distributions (see de
Grijs \& Goodwin 2007). In all panels of Fig. \ref{smcclfs4.fig}, we
have overplotted CMFs with the canonical slope of $\alpha = 2$
(corresponding to a slope of $-1$ in units of ${\rm d} \log(M_{\rm
cl}/{\rm M}_\odot) / {\rm d} \log (N_{\rm cl})$, used in these
panels). We have only shifted and scaled these lines vertically, as
justified below.

We emphasise that we need to choose the age ranges of our cluster
subsamples carefully, for both physical reasons and also because of
the discrete nature of the model isochrones. Regarding the latter, it
is well known that broad-band SED fitting results in artefacts in the
cluster age distribution. This is predominantly caused by specific
features in the SSP models, such as the onset and presence of red
giant branch or asymptotic giant branch (AGB) stars at, respectively,
$\sim 10$ and $\sim 100$ Myr (e.g., Bastian et al. 2005).
Alternatively, both the age-metallicity and the age-extinction
degeneracies will affect the resulting cluster age distributions, thus
also leading to artefacts in the data (e.g., de Grijs et al. 2003;
Anders et al. 2004). We have attempted to avoid placing our age range
boundaries around ages (and, where possible, have taken account of the
uncertainties in age in doing so) where the effects of such artefacts
might seriously impede the interpretation of the results. For
instance, one can see a clear artefact in the cluster age distribution
(which we will refer to as a `chimney') at $\log( t / {\rm yr} )
\simeq 7.2 \, (\simeq 16$ Myr); the average uncertainties for these
ages are of order a few Myr, so that we decided to limit our youngest
cluster subsample to clusters younger than 10 Myr. If, instead, we had
adopted an age limit at $\log( t / {\rm yr} ) = 7.17 \, (15$ Myr), we
would have had marginally better statistics, but our analysis would be
affected by the unknown effects of the age uncertainties associated
with this chimney (see Goodwin et al., in prep., for a detailed
discussion of the issues involved).

The rationale for adopting as our youngest subsample all clusters with
ages $\le 10$ Myr is that at these young ages, the vast majority of
the star clusters present will still be detectable, even in the
presence of early gas expulsion (e.g., Goodwin \& Bastian 2006) -- as
long as they are optically conspicuous. The CMF of this subsample is
shown in Fig. \ref{smcclfs4.fig}a.

Fig. \ref{smcclfs4.fig}b includes our sample clusters with ages in
excess of 40 Myr, up to 160 Myr. While the upper age limit ensures the
full inclusion of the clusters affected by the onset of the AGB stage,
its exact value is rather unimportant for our analysis, and it was
mainly determined by the need to have reasonable statistics in this
and the upper age range, shown in Fig. \ref{smcclfs4.fig}c. The lower
age limit of this subsample is crucial, however. As shown by Goodwin
\& Bastian (2006), most dissolving clusters will have dispersed by an
age of $\sim 30$ Myr, while the surviving clusters will have returned
to an equilibrium state by $\sim 40$ Myr, when some of the early
expansion will have been reversed, depending on the effective
star-formation efficiency. This latter age is therefore a good lower
boundary to assess the surviving star cluster population.

We explicitly exclude any star clusters aged between 10 and 40 Myr
from our analysis. In this age range, it is likely that dissolving
star clusters that will not survive beyond about 30--40 Myr might
still be detectable and therefore possibly contaminate our sample. In
addition, this is the age range in which early gas expulsion causes
rapid cluster expansion, before settling back into equilibrium at
smaller radii; because of the expanded nature of at least part of the
cluster sample, we might not be able to detect some of the
lower-luminosity (and hence lower-mass) clusters that might again show
up beyond an age of $\sim 40$ Myr. At the same time, the effects of
`infant weightloss' (Weidner et al. 2007) will further confuse the
analysis in this age range.

The scaled canonical CMF in Fig. \ref{smcclfs4.fig}b is an almost
perfect fit to the observed CMF. The best-fitting CMF slope
is ${\rm d} \log(M_{\rm cl}/{\rm M}_\odot) / {\rm d} \log (N_{\rm cl})
= -0.82 \pm 0.18$, but this compares to ${\rm d} \log(M_{\rm cl}/{\rm
M}_\odot) / {\rm d} \log (N_{\rm cl}) = -1.01 \pm 0.20$ if we ignore
the lowest-mass clusters at $\log(M_{\rm cl}/{\rm M}_\odot) \le 3.2$,
where there may be residual incompleteness effects.

This very good match between the observed CMF for the age range from
40--160 Myr (Fig. \ref{smcclfs4.fig}b) and the scaled CMF from
Fig. \ref{smcclfs4.fig}a implies that {\it the SMC cluster system has
not been affected by any significant amount of cluster infant
mortality for cluster masses greater than a few} $\times 10^3$
M$_\odot$. Based on a detailed assessment of the uncertainties in both
the CMFs and the age range covered by our youngest subsample, we can
limit the extent of infant mortality between the youngest and the
intermediate age range to a maximum of $\lesssim 30$ per cent
($1\sigma$). We rule out a $\sim 90$ per cent mortality rate per
decade of age at a $>6 \sigma$ level. This result is in excellent
agreement with that of Gieles et al. (2007) -- although we also note
that Chandar et al. (2006) do not include the youngest SMC clusters in
their analysis. Using the age distribution of the SMC cluster sample
in units of the number of clusters observed per unit time-scale, we
independently confirm this scenario (de Grijs \& Goodwin 2007).

\section{Dodgy diagnostics?}
\label{diagn.sec}

\begin{figure}[t]
\includegraphics[width=\columnwidth]{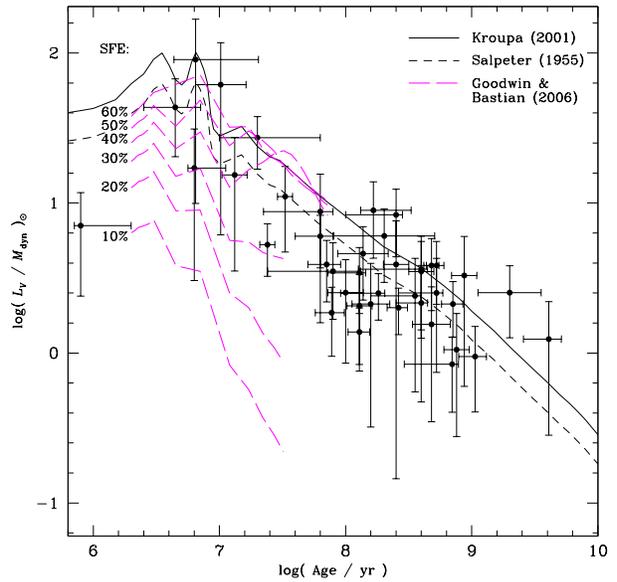}
\caption{\label{ocl.fig}Diagnostic age versus $M/L$ ratio diagram,
including the Galactic open clusters for which velocity dispersion
measurements are available (adapted from de Grijs et al. 2008); the
figure coding is as for Fig. \ref{diagnostic.fig} }
\label{fig:2}
\end{figure}

We now return to the use of the $M/L$ ratio versus age diagram as a
diagnostic tool. Despite the myriad uncertainties associated with its
use, using this approach one can get at least an initial assessment
as to whether a given cluster may be (i) significantly out of virial
equilibrium, in particular `super-virial', or (ii) significantly
overabundant in low-mass stars. Since the ground-breaking work by
Bastian \& Goodwin (2006) and Goodwin \& Bastian (2006), we can now
also model any (super-virial) deviations from the SSP models for the
youngest ages, if we assume that these are predominantly a function of
the effective SFE.

This has led a number of authors to suggest that, in the absence of
significant external perturbations, massive clusters located in the
vicinity of the SSP models and aged $\ga 10^8$ yr may survive for a
Hubble time and eventually become old GC-like objects (e.g., Larsen et
al. 2004; Bastian et al. 2006; de Grijs \& Parmentier 2007).

Encouraged by the recent progress in this area based on both
observational and theoretical advances, in de Grijs et al. (2008) we
explore whether we can also use the same diagnostic diagram to assess
the stability, formation conditions and longevity of those open
clusters in the Milky Way for which the required observational data
exist in the literature, and hence whether this approach might be
useful in view of future data mining opportunities.

Using a sample of Galactic open clusters for which reasonably accurate
internal velocity dispersions are available in the literature, we
constructed a homoge\-nised set of observational data drawn from a wide
variety of publications, also including their most likely uncertainty
ranges. This allowed us to derive dynamical mass estimates for our
sample of open clusters, as well as their respective $M/L$ ratios and
-- crucially -- the associated (realistic) uncertainties.

Although our sample of Galactic open clusters is by no means
statistically complete in any sense, this study has provided useful
additional constraints on the dynamical state of the individual sample
clusters. Compared to the photometric evolution predicted on the basis
of Salpeter (1955) or Kroupa (2001)-type stellar IMFs, this allowed us
to independently assess the clusters' stability with respect to
internal dynamical effects (and -- to some extent -- also to external
perturbations).

Most importantly, we conclude that for an open cluster to survive for
any significant length of time (in the absence of substantial external
perturbations), it is a necessary but not a sufficient condition to be
located close to the predicted photometric evolutionary sequences for
`normal' SSPs. This is highlighted using a number of our sample
clusters which are known to be in a late stage of dissolution, yet lie
very close indeed to either of the evolutionary sequences defined by
the Salpeter (1955) or Kroupa (2001) IMFs.  However, we also note that
a significant fraction of our sample clusters show the signatures of
dynamical relaxation and stability. Despite their relatively small
masses ($M_{\rm cl} \la 2 \times 10^3$ M$_\odot$) and ages in excess
of a few $\times 10^8$ yr, this is not unexpected. Using the vertical
oscillation period, $\pi$, around the Galactic plane of NGC 2323 ($\pi
\approx 50$ Myr; Clari\'a, Piatti \& Lapasset 1998) as an example,
this cluster has only been through a few of these periods, given its
age of $\log (t / {\rm yr}) = 8.11^{+0.05}_{-0.25}$ (Kalirai et
al. 2003). However, at the Galactocentric distance of the Sun, a
Pleiades-like open cluster crosses the Galactic disc approximately
10--20 times before it dissolves (de la Fuente Marcos 1998a,b).

Finally, we caution that for the low-mass Galactic open clusters in
particular, the measured velocity dispersions may be significantly
affected by the orbital motions of a sizeable fraction of binary or
multiple systems (e.g., Kouwenhoven \& de Grijs 2008). In a follow-up
paper (Kouwenhoven et al., in prep.) we will explore this
quantitatively using $N$-body simulations.

\section{`Super' star cluster survival confirmed?}

We recently reported the discovery of an extremely massive, but old
($12.4 \pm 3.2$ Gyr) GC in M31, 037-B327, that has all the
characteristics of having been an exemplary YMC at earlier times,
based on an extrapolation of its present-day extinction-corrected
$V$-band luminosity back to an age of 10 Myr (Ma et al. 2006b; see
also Cohen 2006). To have survived for a Hubble time, we concluded
that its stellar IMF cannot have been top-heavy. Using this
constraint, and a variety of SSP models, we determined a {\it
photometric} mass for 037-B327 of $M_{\rm GC} = (3.0 \pm 0.5)\times
10^7$ M$_\odot$, somewhat depending on the SSP models used, the
metallicity and age adopted and the IMF representation. In view of the
large number of free parameters, the uncertainty in our photometric
mass estimate is surprisingly small (although this was recently
challenged by Cohen 2006). This mass, and its relatively small
uncertainties, make this object potentially one of the most massive
star clusters of any age in the Local Group. Based on a more recent
dynamical mass determinations by Cohen (2006), it appears that
037-B327 may be a factor of $\sim 2-3$ less massive than M31 G1,
assuming that both GCs have the same stellar IMF. Nevertheless, this
still confirms the nature of 037-B327 as one of the most massive star
clusters in the Local Group. As a surviving `super' star cluster, this
object is therefore of prime importance for theories aimed at
describing massive star cluster evolution.

Cohen (2006) suggests that the high mass estimate of Ma et al. (2006b)
may have been affected by a non-uniform extinction distribution across
the face of the cluster (see also Ma et al. 2006a for a more detailed
discussion). She obtains, from new $K$-band imaging and different
assumptions on the extinction affecting the $K$-band light, that $M_K$
of 037-B327 may be some 0.16 mag brighter than that of M31 G1, or
about twice as luminous. Despite these corrections provided by Cohen
(2006), the basic conclusion from Ma et al. (2006b), i.e., that at the
young age of 10 Myr cluster M31 037-B327 must have been a benchmark
example of a `super' star cluster, and that its IMF must thus have
contained a significant fraction of low-mass stars, still stands
firmly.

Thus, in summary, the formation of GCs, which was once thought to be
limited to the earliest phases of galaxy formation, appears to be
continuing at the present time in starburst, interacting and merging
galaxies in the form of star clusters with masses and compactnesses
typical of GCs. Whether these YMCs will evolve to become old GCs by
the time they reach an age of 13 Gyr depends to a very large extent on
their environment, however. For a host galaxy with a smooth
logarithmic gravitational potential, the ambient density seems to be
the key parameter driving the rate of cluster evolution. This is
accelerated in the presence of substructure in the host galaxy, such
as that commonly provided by bulge, spiral arm and giant molecular
cloud components (see also the review of de Grijs \& Parmentier 2007,
and references therein).

%
\acknowledgments I am grateful to Genevi\`eve Parmentier, Simon
Goodwin, Pavel Kroupa and Jun Ma for discussions and collaborative
work on which most of this contribution is based. I would also like to
express my thanks to Enrique P\'erez and Rosa Gonz\'alez Delgado (and
by extension to the Local Organising Committee) for their significant
efforts to make this conference proceed as smoothly as possible, and
to the members of the Scientific Organising Committee for their expert
advice. I acknowledge funding from the Royal Society allowing me to
attend this conference.


%

%

\end{document}